\documentclass[preprint]{ptephy_v2}

\usepackage{hyperref}

\begin{document}

\title{Constraints on Fermionic Dark Matter Absorption from Radiochemical Solar-Neutrino Measurements}

\author[1]{K.~Ishidoshiro}
\author[1]{K.~Tachibana}%

\affil[1]{Research Center for Neutrino Science, Tohoku University, Sendai 980-8578, Japan}%




\begin{abstract}
We reinterpret classic radiochemical solar-neutrino measurements as ``rate meters'' for additional,
non-negative capture-like contributions induced by fermionic dark matter absorption.
Using the chlorine and gallium production-rate data, we build a Bayesian likelihood that accounts for the dominant uncertainties in the solar-neutrino capture-rate prediction (solar fluxes, oscillation parameters, and capture cross sections). 
Solar-model metallicity systematics are made explicit by presenting results for both the B16--GS98
and B16--AGSS09met solar-model realizations. 
From the 1D marginalized posteriors of the joint $(R_{\chi,\mathrm{Cl}},R_{\chi,\mathrm{Ga}})$ analysis,
we obtain 90\% upper limits on additional capture-like rate contributions, dominated by chlorine:
$R_{\chi,\mathrm{Cl},90}\simeq 0.388~\mathrm{SNU}$ (B16--GS98) and $0.588~\mathrm{SNU}$ (B16--AGSS09met). 
In the charged-current V--A benchmark, we map these constraints onto upper bounds on $y\equiv m_\chi^2/(4\pi\Lambda^4)$ for $m_\chi$ above the ${}^{71}$Ga and ${}^{37}$Cl capture thresholds, using a pep-normalized operator mapping anchored to solar-neutrino capture inputs, where $m_\chi$ is the dark matter mass and $\Lambda$ is the effective scale suppressing the charged-current operator.
At $m_\chi\simeq 1~\mathrm{MeV}$, we find $y_{90}\simeq 4.88\times 10^{-49}~\mathrm{cm}^2$
(B16--GS98) and $7.08\times 10^{-49}~\mathrm{cm}^2$ (B16--AGSS09met).
These radiochemical bounds are complementary to xenon-based absorption searches and collider
interpretations by probing distinct nuclear targets with minimal reliance on spectral reconstruction.
\end{abstract}


\maketitle

\section{Introduction}
\label{sec:intro}

The particle nature of dark matter (DM) remains unknown despite compelling gravitational evidence across astrophysical and cosmological scales.
A broad experimental program has therefore been developed to search for non-gravitational interactions of DM with Standard Model (SM) particles.
Conventional direct-detection experiments predominantly target elastic scattering off electrons or nuclei, but their sensitivity can be strongly degraded for sub-GeV DM by recoil kinematics and detector thresholds.
These considerations motivate qualitatively different detection modes in which the incoming DM particle is \emph{absorbed} rather than scattered, so that its rest-mass energy participates directly in the final-state kinematics.

In this context, absorption of light fermionic DM provides a well-motivated and experimentally distinctive target.
An effective-operator description of fermionic absorption on nuclear targets has been developed in Refs.~\citep{Dror:2020FDM_PRL,Dror:2020FDM_JHEP}, which classifies interactions into neutral-current- and charged-current-like structures and delineates the associated experimental signatures. 
Because the absorption rate is proportional to the DM number density, absorption searches are naturally sensitive to light DM masses. 
In the charged-current-like case, precision measurements of solar-neutrino capture rates provide a powerful handle. 
More broadly, absorption leads to a characteristic energy deposition, so complementary constraints can be obtained from direct-detection searches for absorption signals on nuclear and electron targets~\citep{Gu:2022PandaX4T,Arnquist:2024MajoranaDM,Dai:2022CDEX10,Adams:2025PICO60,PandaX:2022ood,Ge:2022ius}.

In this work we derive constraints on fermionic DM absorption by using radiochemical solar-neutrino measurements as rate meters for additional charged-current-like contributions to the daughter-nucleus production rate. 
Radiochemical experiments such as the chlorine and gallium detectors measure time-integrated production rates of daughter nuclei.
Any absorption-induced capture channel would contribute a strictly non-negative rate on top of the solar-neutrino capture-rate prediction, enabling existing radiochemical datasets to be reinterpreted as limits on DM absorption with minimal reliance on detector-specific spectral modeling or event-by-event reconstruction. 

It is worth noting that the observed chlorine and gallium radiochemical production rates are broadly consistent with the corresponding solar-neutrino capture-rate  expectations once the electron-neutrino survival probability is evaluated using oscillation parameters constrained independently by reactor antineutrino experiments. This agreement implies that only limited room remains for additional non-negative capture-like contributions, which naturally motivates reinterpreting the radiochemical datasets as sensitive probes of fermionic DM absorption.

A central ingredient in this reinterpretation is the prediction of the solar-neutrino capture rates, which depends on solar-neutrino fluxes, neutrino flavor conversion, and nuclear/atomic inputs entering the capture cross sections.
We therefore construct a Bayesian likelihood that combines the radiochemical production-rate measurements with priors on oscillation parameters and a detailed treatment of flux and capture cross-section normalization systematics, implemented through the corresponding nuisance parameters.
To capture the dominant solar-model dependence, we treat the spread between representative high- and low-metallicity standard solar models (SSMs) as an explicit theory systematic, and present results for both B16--GS98 and B16--AGSS09met SSM realizations.

To enable a direct comparison with existing absorption searches and collider limits on the same effective operators, we adopt the charged-current V--A operator of Refs.~\citep{Dror:2020FDM_PRL,Dror:2020FDM_JHEP} as our benchmark choice and translate our rate limits onto the standard coupling combination
\begin{equation}
y \equiv \frac{m_\chi^2}{4\pi\Lambda^4},
\end{equation}
which has the dimensions of a cross section and is widely used to report constraints on charged-current fermionic absorption. $m_\chi$ is the DM mass and $\Lambda$ denotes the effective scale suppressing the relevant operator. Hereafter, we refer to this benchmark simply as charged-current absorption unless otherwise noted.

A distinctive feature of our analysis is a data-driven operator mapping:
for each target, we calibrate the dominant nuclear response to the tabulated pep $\nu_e$ capture cross section of the same nucleus, thereby absorbing the leading target-dependent nuclear-structure uncertainty into a single effective matrix element. This ``pep-normalized'' mapping provides a robust anchor that is consistent with the capture inputs used in the rate prediction.

We first present a model-independent rate-level constraint to isolate what is purely data-driven, and then interpret it within a specific charged-current absorption benchmark. 
This paper is organized as follows.
Sec.~\ref{sec:inputs} describes the radiochemical datasets and the analysis inputs used to predict the solar-neutrino capture rates.
In Sec.~\ref{sec:stat} we present the likelihood construction and the treatment of systematic uncertainties, including the operator mapping used to report limits in terms of $y$.
Results are shown in Sec.~\ref{sec:results}.
We discuss dominant systematics and compare the results to existing constraints in Sec.~\ref{sec:discussion}, and summarize our conclusions in Sec.~\ref{sec:conclusion}.

\section{Analysis Inputs}
\label{sec:inputs}

Our analysis combines radiochemical production-rate measurements with the theoretical inputs needed to predict the solar-neutrino capture contribution.
All rates are expressed in solar neutrino units (SNU), where $1~\mathrm{SNU}\equiv 10^{-36}$ captures per target atom per second.
Below we summarize the datasets and the main analysis inputs.

We use the time-integrated radiochemical production rates measured in the chlorine and gallium programs.
For chlorine, we adopt the final Homestake ${}^{37}$Cl result,
$R_{\mathrm{Cl}}^{\mathrm{obs}} = 2.56 \pm 0.16~(\mathrm{stat}) \pm 0.16~(\mathrm{syst})~\mathrm{SNU}$~\citep{Cleveland:1998}.
We combine the statistical and systematic uncertainties in quadrature, yielding
$R_{\mathrm{Cl}}^{\mathrm{obs}} = 2.56 \pm 0.23~\mathrm{SNU}$.
For gallium, we use as our default input the weighted average of SAGE, GALLEX, and GNO,
$R_{\mathrm{Ga}}^{\mathrm{obs}} = 66.1 \pm 3.1~\mathrm{SNU}$~\citep{Altmann:2005,Abdurashitov:2009}.

These measurements are sensitive to charged-current $\nu_e$ capture on the relevant targets,
\begin{equation}
\nu_e + (A, Z) \to e^- + (A, Z + 1)\,,
\end{equation}
and therefore constrain any additional non-negative capture-like contribution once the reaction is kinematically accessible.
The kinematic thresholds are $E_{\nu,\mathrm{Cl}}^{\rm th}=0.814~\mathrm{MeV}$ and $E_{\nu,\mathrm{Ga}}^{\rm th}=0.233~\mathrm{MeV}$~\citep{Bahcall:1997GaXS,Bahcall:1998Uncertainties},
making gallium sensitive to the low-energy pp-dominated flux in addition to higher-energy components.

The predicted solar-neutrino capture rate depends on the solar-neutrino fluxes $\Phi_k$ from the dominant production channels
$k=\{\mathrm{pp},\mathrm{pep},{}^{7}\mathrm{Be},{}^{8}\mathrm{B},{}^{13}\mathrm{N},{}^{15}\mathrm{O},{}^{17}\mathrm{F},\mathrm{hep}\}$.
The SSM provides self-consistent predictions for these fluxes and their theoretical uncertainties; we adopt the B16 SSM central values and quoted uncertainties~\citep{B16SSM}.
To make the dominant solar-model systematic explicit, we bracket the metallicity dependence using the B16--GS98~\citep{GS98} and B16--AGSS09met~\citep{AGSS09} realizations. 
Here, B16-GS98 and B16-AGSS09met correspond to the high- and
low-metallicity realizations of the SSM, respectively. The difference arises from the heavy-element abundances adopted in the solar composition, which affect the solar interior opacity and therefore the predicted solar-neutrino fluxes.
In practice, the higher-metallicity B16-GS98 model generally predicts slightly larger solar-neutrino capture rates for the radiochemical targets considered here. Consequently, less room remains for an additional non-negative DM-induced contribution, which leads to stronger limits than those obtained with the lower-metallicity B16-AGSS09met model.

For each target $T\in\{\mathrm{Cl},\mathrm{Ga}\}$, the capture-rate prediction depends on the charged-current capture cross section $\sigma_T(E)$.
We adopt $\sigma_T(E)$ for ${}^{37}$Cl and ${}^{71}$Ga (with quoted uncertainties) from the Bahcall \emph{et al.}\ calculations and subsequent updates~\citep{Bahcall:1996B8Spec,Bahcall:1997GaXS,Bahcall:1998Uncertainties}, and evaluate the individual rate contributions by numerical integration over the standard solar spectra.

The prediction for each radiochemical rate depends on the electron-neutrino survival probability $P_{ee}(E)$ through
\begin{align}
R_{\nu,k,T} &= 10^{36}\,\Phi_k\int dE\,\phi_k(E)\,\sigma_T(E)\,P_{ee}(E)\,,\\
R_{\nu,T} &= \sum_k R_{\nu,k,T}\,,
\label{eq:Rnu_general}
\end{align}
where $\phi_k(E)$ is the normalized energy spectrum of component $k$ (including line spectra for monochromatic sources).
We evaluate $P_{ee}(E)$ in a three-flavor Mikheyev--Smirnov--Wolfenstein (MSW) framework using an adiabatic approximation~\citep{Wolfenstein:1978,MSW:1985,Parke:1986,KuoPantaleone:1989}.

We incorporate external constraints on the oscillation parameters entering $P_{ee}(E)$.
To avoid potential circularity with solar-neutrino rate information, we take the ``solar'' parameters $(\sin^2\theta_{12},\Delta m_{21}^2)$ from reactor $\bar\nu_e$ oscillation measurements.
Concretely, we adopt the first JUNO reactor result (Normal Ordering)~\citep{JUNOFirstOsc}:
$\sin^2\theta_{12} = 0.3092 \pm 0.0087$ and
$\Delta m_{21}^2 = (7.50 \pm 0.12)\times 10^{-5}\ \mathrm{eV}^2$.
For $\theta_{13}$ we use the Particle Data Group (PDG) world average~\citep{PDG2024}:
$\sin^2\theta_{13} = (2.19 \pm 0.07)\times 10^{-2}$.
For solar $\nu_e$ survival probabilities, $\theta_{23}$ and $\delta_{\rm CP}$ do not enter at leading order and are neglected.

\section{Analysis Framework}
\label{sec:stat}

We infer upper bounds on additional, non-negative capture-like contributions by comparing radiochemical production-rate measurements to the solar $\nu_e$ capture.
The capture-rate prediction depends on nuisance parameters describing solar-neutrino fluxes, neutrino oscillations, and capture cross-section inputs.
We treat the DM-induced contribution either (i) phenomenologically as independent additive rates (in SNU) for each target, or (ii) in our charged-current absorption interpretation, parameterized by
$y \equiv m_\chi^2/(4\pi\Lambda^4)$~\citep{Dror:2020FDM_PRL,Dror:2020FDM_JHEP}.
Throughout, we adopt flat priors on the non-negative DM signal parameters and compute one-sided Bayesian upper limits at 90\% credible level.

For each target $T\in\{\mathrm{Cl},\mathrm{Ga}\}$, the predicted total radiochemical production rate is written as
\begin{equation}
R_T^{\mathrm{pred}}(\boldsymbol{\eta},R_{\chi,T})
= R_{\nu,T}(\boldsymbol{\eta}) + R_{\chi,T}\,,
\label{eq:rate_model}
\end{equation}
where $\boldsymbol{\eta}$ collects nuisance parameters and $R_{\chi,T}\ge 0$ denotes the DM-induced capture-like contribution on target $T$.

Within a fixed SSM realization, we model flux uncertainties as multiplicative rescalings,
\begin{equation}
\Phi^{\rm SSM}_k\  \to\  \Phi_k^{\rm SSM}\,\xi_k\,,
\label{eq:flux_prior_inputs}
\end{equation}
where $\Phi_k^{\rm SSM}$ is the SSM central value.
We assign independent lognormal priors to the flux scalings,
$\ln \xi_k \sim \mathcal{N}(-\sigma_{\Phi_k}^2/2,\sigma_{\Phi_k}^2)$,
which reproduce the quoted fractional uncertainties to leading order, where $\sigma_{\Phi_k}$ is the quoted fractional $1\sigma$ uncertainty for component $k$.

To propagate capture cross-section systematics in a compact way, we introduce a target-dependent normalization parameter $\alpha_T$ that rescales the overall capture-rate prediction for that target,
\begin{equation}
R_{\nu,T}\ \to\ \alpha_T\,R_{\nu,T}\,,
\label{eq:alpha_scaling_inputs}
\end{equation}
with $\ln \alpha_T \sim \mathcal{N}(-\delta_T^2/2,\delta_T^2)$.
Here $\delta_T=\sqrt{\sum_k \omega^2_{k,T}\delta^2_{k,T}}$ is an effective fractional normalization uncertainty for target $T$ obtained by compressing the quoted per-component capture cross-section uncertainties into a single parameter, with $\omega_{k,T}=R_{0,k,T}/\sum_j R_{0,j,T}$.
$R_{0,k,T}$ denotes the nominal unoscillated contribution of component $k$.
We use the same $\alpha_T$ in the operator mapping below, so that a single target-level normalization captures the leading uncertainty in the dominant allowed strength affecting both the capture prediction and the absorption mapping.

To capture residual uncertainty in the MSW matter potential within the adiabatic approximation, we introduce a single nuisance parameter that rescales the solar electron density entering the potential.
In the adiabatic treatment, the survival probability depends on the local electron density at the neutrino production point; for each solar flux component $k$ we therefore evaluate the matter potential at a representative production radius $r_k^{\rm eff}$ defined as the production-weighted mean radius for component $k$,
\begin{equation}
V_e(r_k^{\rm eff})=\sqrt{2}\,G_F\,N_e(r_k^{\rm eff})\,,
\end{equation}
and encode residual uncertainty by an overall rescaling,
\begin{equation}
N_e(r_k^{\rm eff}) \to (1+\eta_{N_e})\,N_e(r_k^{\rm eff}),\qquad
\eta_{N_e}\sim \mathcal{N}(0,\sigma_{N_e}^2)\,.
\label{eq:etaV_inputs}
\end{equation}
We adopt $\sigma_{N_e}=0.10$, which conservatively parametrizes model-to-model differences in the solar electron-density profile as well as the effective-radius approximation.

In the rate-level (model-independent) parameterization, we treat the DM contribution as an independent non-negative additive rate in SNU,
\begin{equation}
R_{\chi,T} = R_{\chi,T}^{\mathrm{(free)}}\ge 0\,.
\label{eq:Rchi_free}
\end{equation}
The superscript “(free)” indicates that these parameters are treated
as model-independent free variables representing non-negative
additive capture-like rates expressed in SNU, prior to imposing
the charged-current absorption mapping used in the benchmark
operator interpretation discussed below.

In the charged-current absorption,  we instead relate the capture-like rate to the coupling parameter $\kappa\equiv \Lambda^{-4}$ via
\begin{equation}
R_{\chi,T}(\boldsymbol{\eta},\kappa;m_\chi) = 10^{36}\frac{\rho_\chi}{2m_\chi}\,\alpha_T \, A_T(m_\chi)\,\kappa \,,
\label{eq:Rchi_ccva}
\end{equation}
where $\rho_\chi=0.3$~GeV/cm$^3$ is the local DM energy density.
The response factor $A_T(m_\chi)$ collects the allowed kinematics and an effective nuclear matrix element in the charged-current absorption,
\begin{equation}
A_T(m_\chi) \equiv B_0(E_e)\,M_{\rm eff,T}^2,
\label{eq:AT_def}
\end{equation}
with the allowed kinematic factor
\begin{equation}
B_0(E_e)= \frac{p_e E_e}{\pi}\,F(Z+1,E_e)\,,
\label{eq:B0_def}
\end{equation}
where $p_e=\sqrt{E_e^2-m_e^2}$ and $E_e=m_e+(m_\chi-E_{\nu,T}^{\rm th})$.
The factor $1/2$ in Eq.~\eqref{eq:Rchi_ccva} follows the convention of Refs.~\citep{Dror:2020FDM_PRL,Dror:2020FDM_JHEP}.

For $M_{\rm eff,T}$ we adopt a pep-normalized prescription in which nuclear-structure dependence is absorbed into a single effective matrix element by matching, for each target, the tabulated pep $\nu_e$ capture cross section at the line energy.
Specifically, using the pep capture cross section $\sigma_{\rm pep,T}\equiv \sigma_{\nu_e T}(E_{\rm pep})$, we define
\begin{equation}
M_{\mathrm{eff},T}^2
\equiv
\frac{\sigma_{\mathrm{pep},T}}{B_0(E_{e,\mathrm{pep},T})},
\qquad
E_{e,\mathrm{pep},T}= m_e + (E_{\mathrm{pep}}-E_{\nu,T}^{\rm th}),
\label{eq:Meff_from_pep}
\end{equation}
so that
\begin{equation}
A_T(m_\chi)
=
\sigma_{\mathrm{pep},T}\,
\frac{B_0(E_e)}{B_0(E_{e,\mathrm{pep},T})}.
\label{eq:AT_pep_normalized}
\end{equation}
We then translate an upper limit on an additional capture-like rate into a bound on
$y(m_\chi)=m_\chi^2/(4\pi\Lambda^4)=\kappa\,m_\chi^2/(4\pi)$, which has dimensions of area and is quoted in cm$^2$.

For each target $T$, the radiochemical measurement provides an observed rate $R_T^{\rm obs}$ with uncertainty $\delta R_T$.
We adopt a Gaussian likelihood,
\begin{equation}
\mathcal{L}_T(R_{\chi,T}^{\mathrm{(free)}},\boldsymbol{\eta}) \propto
\exp\!\left[
-\frac{1}{2}\left(\frac{R_T^{\mathrm{obs}} - R_T^{\mathrm{pred}}(\boldsymbol{\eta},R_{\chi,T}^{\mathrm{(free)}})}{\delta R_T}\right)^2
\right],
\label{eq:LT}
\end{equation}
and assume the chlorine and gallium measurements are statistically independent, so that
\begin{equation}
\mathcal{L}_{\mathrm{comb}}(R_{\chi,\mathrm{Cl}}^{\mathrm{(free)}},R_{\chi,\mathrm{Ga}}^{\mathrm{(free)}},\boldsymbol{\eta})
=
\mathcal{L}_{\mathrm{Cl}}(R_{\chi,\mathrm{Cl}}^{\mathrm{(free)}},\boldsymbol{\eta})\,
\mathcal{L}_{\mathrm{Ga}}(R_{\chi,\mathrm{Ga}}^{\mathrm{(free)}},\boldsymbol{\eta})\,.
\label{eq:Lcomb}
\end{equation}
For the charged-current absorption at fixed $m_\chi$, we evaluate the same combined likelihood with $R_{\chi,T}$ set by Eq.~\eqref{eq:Rchi_ccva}, yielding $\mathcal{L}_{\mathrm{comb}}(\kappa,\boldsymbol{\eta};m_\chi)$.

We assign priors as follows:
(i) flux scalings $\xi_k$ are independent lognormal factors with mean unity and widths given by the quoted fractional flux uncertainties;
(ii) oscillation parameters $(\sin^2\theta_{12},\Delta m_{21}^2,\sin^2\theta_{13})$ have Gaussian priors based on external constraints restricted to physical domains;
(iii) the capture cross-section normalization parameters $\alpha_T$ follow the lognormal prior above;
and (iv) the matter-potential (electron-density) scaling $\eta_{N_e}$ has a Gaussian prior.
For the DM signal parameters we adopt flat priors for $R_{\chi,T}^{\mathrm{(free)}}\ge 0$ (rate-level analysis) and for $\kappa\ge 0$ (charged-current absorption).

We marginalize the combined likelihood over nuisance parameters describing flux, oscillation, and capture-input uncertainties.
For the rate-level analysis we compute
\begin{equation}
\mathcal{L}_{\mathrm{marg}}(R_{\chi,\mathrm{Cl}}^{\mathrm{(free)}},R_{\chi,\mathrm{Ga}}^{\mathrm{(free)}})=
\int d\boldsymbol{\eta}\;\mathcal{L}_{\mathrm{comb}}(R_{\chi,\mathrm{Cl}}^{\mathrm{(free)}},R_{\chi,\mathrm{Ga}}^{\mathrm{(free)}},\boldsymbol{\eta})\,\pi(\boldsymbol{\eta})\,,
\label{eq:Lmarg}
\end{equation}
and the corresponding posterior
\begin{equation}
p(R_{\chi,\mathrm{Cl}}^{\mathrm{(free)}},R_{\chi,\mathrm{Ga}}^{\mathrm{(free)}}\,|\,\mathrm{data})
\propto
\mathcal{L}_{\mathrm{marg}}(R_{\chi,\mathrm{Cl}}^{\mathrm{(free)}},R_{\chi,\mathrm{Ga}}^{\mathrm{(free)}})\,
\pi(R_{\chi,\mathrm{Cl}}^{\mathrm{(free)}})\,\pi(R_{\chi,\mathrm{Ga}}^{\mathrm{(free)}})\,,
\label{eq:posterior}
\end{equation}
where $\pi$ denotes the prior probability density.
In particular, $\pi(\boldsymbol{\eta})$ represents the joint prior for the nuisance parameters included in the analysis, while $\pi(R_{\chi,T}^{\mathrm{(free)}})$  denotes the prior for the dark-matter-induced capture-like rate parameters with $T\in\{\mathrm{Cl},\mathrm{Ga}\}$. 
We obtain one-sided 90\% credible upper limits from the 1D marginalized posteriors, e.g.
\begin{equation}
\int_{0}^{R_{\chi,T,90}^{\mathrm{(free)}}} dR_{\chi,T}^{\mathrm{(free)}}\;
p(R_{\chi,T}^{\mathrm{(free)}}\,|\,\mathrm{data}) = 0.9\,,
\label{eq:limit90_rate}
\end{equation}

For the charged-current absorption at fixed $m_\chi$, we marginalize over nuisance parameters to obtain $\mathcal{L}_{\mathrm{marg}}(\kappa;m_\chi)$ and the posterior $p(\kappa\,|\,\mathrm{data},m_\chi)$, from which we define $\kappa_{90}(m_\chi)$ by
\begin{equation}
\int_{0}^{\kappa_{90}(m_\chi)} d\kappa\;
p(\kappa\,|\,\mathrm{data},m_\chi) = 0.9\,,
\label{eq:limit90_kappa}
\end{equation}
and translate to $y_{90}(m_\chi)$.

\section{Results}
\label{sec:results}

In this section we present constraints on additional non-negative capture-like contributions from fermionic DM absorption using radiochemical solar-neutrino measurements.
We first report rate-level (model-independent) upper limits on additive contributions in SNU, and then present our main charged-current limits, in terms of
$y = m_\chi^2/(4\pi\Lambda^4)=\kappa m_\chi^2/(4\pi)$~\citep{Dror:2020FDM_PRL,Dror:2020FDM_JHEP}.
Results are shown for both B16 SSM realizations (GS98 and AGSS09met), which bracket the dominant metallicity-driven flux systematic. 

\subsection{Phenomenological capture-rate limits}
\label{subsec:results_Rchi}

\begin{figure*}[t]
\centering
\includegraphics[width=0.86\textwidth]{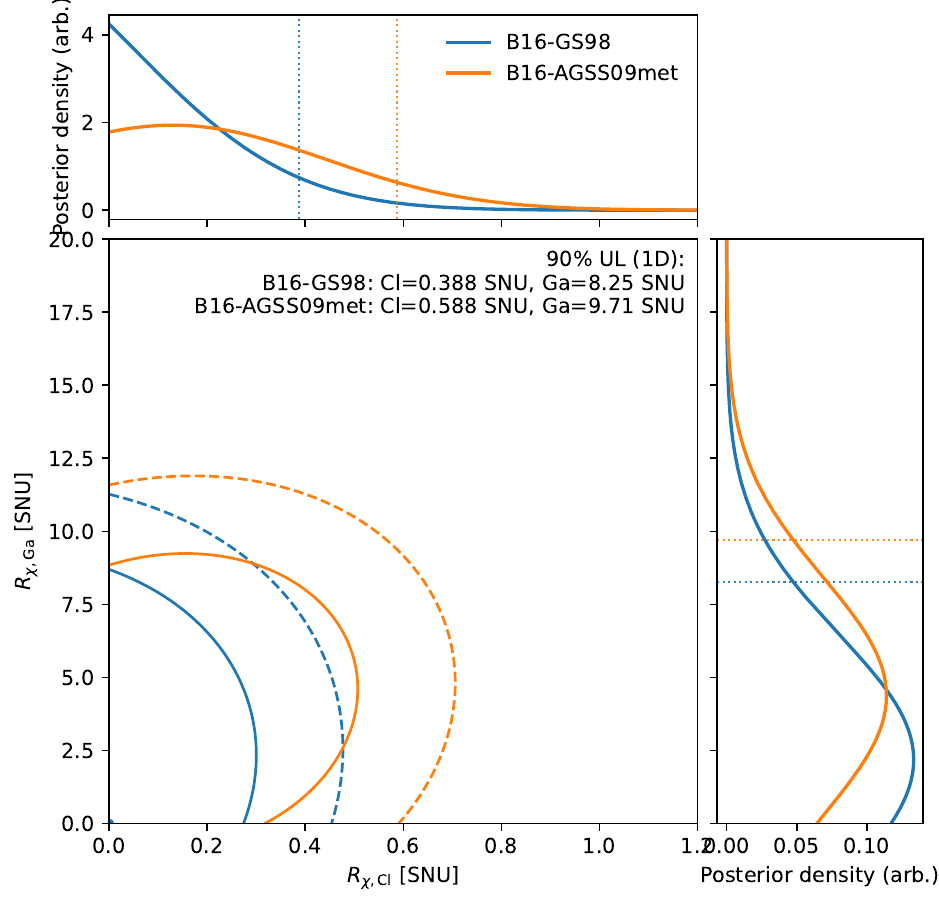}
\caption{
Posterior constraints on independent additive DM-induced rates $(R_{\chi,\mathrm{Cl}}^{\mathrm{(free)}},R_{\chi,\mathrm{Ga}}^{\mathrm{(free)}})$ for the two B16 SSM realizations.
The central panel shows 68\% (inner, solid) and 90\% (outer, dashed) credible-region contours in the $(R_{\chi,\mathrm{Cl}}^{\mathrm{(free)}},R_{\chi,\mathrm{Ga}}^{\mathrm{(free)}})$ plane, while the top and right panels show the corresponding 1D marginalized posteriors.
Vertical (horizontal) dotted lines indicate the one-sided 90\% credible upper limits on $R_{\chi,\mathrm{Cl}}$ ($R_{\chi,\mathrm{Ga}}$).
}
\label{fig:Rchi_corner}
\end{figure*}

As a rate-level (model-independent) result, we constrain independent additive capture-like rates
$R_{\chi,\mathrm{Cl}}^{\mathrm{(free)}}$ and $R_{\chi,\mathrm{Ga}}^{\mathrm{(free)}}$ (SNU) using the combined chlorine-and-gallium likelihood described in Sec.~\ref{sec:stat}.
Figure~\ref{fig:Rchi_corner} shows the joint posterior for $(R_{\chi,\mathrm{Cl}}^{\mathrm{(free)}},R_{\chi,\mathrm{Ga}}^{\mathrm{(free)}})$, and Table~\ref{tab:Rchi_limits} reports the corresponding one-sided 90\% credible upper limits obtained from the 1D marginalized posteriors.

Because the chlorine measurement has the smallest fractional uncertainty, the joint posterior is
sharply constrained in the $R_{\chi,{\rm Cl}}^{\rm (free)}$ direction, while remaining comparatively weak
in $R_{\chi,{\rm Ga}}^{\rm (free)}$; this behavior is visible in Fig.~\ref{fig:Rchi_corner}.

\begin{table}[t]
\centering
\caption{One-sided 90\% credible upper limits on the additive DM-induced rates
$R_{\chi,\mathrm{Cl}}^{\mathrm{(free)}}$ and $R_{\chi,\mathrm{Ga}}^{\mathrm{(free)}}$ (SNU),
obtained from the 1D marginalized posteriors of the combined chlorine-and-gallium analysis shown in Fig.~\ref{fig:Rchi_corner}.}
\label{tab:Rchi_limits}
\begin{tabular}{lcc}
 & $R_{\chi,\mathrm{Cl},90}^{\mathrm{(free)}}$ & $R_{\chi,\mathrm{Ga},90}^{\mathrm{(free)}}$ \\
\hline
B16--GS98      & 0.388 & 8.25 \\
B16--AGSS09met & 0.588 & 9.71 \\
\end{tabular}
\end{table}

\subsection{Main limits in charged-current absorption}
\label{subsec:results_y}

We next present limits on $y(m_\chi)$ obtained from the combined likelihood in Sec.~\ref{sec:stat}.
At each $m_\chi$ we determine $\kappa_{90}(m_\chi)$ via Eq.~\eqref{eq:limit90_kappa} and convert it to $y_{90}(m_\chi)$.

\begin{figure*}[t]
\centering
\includegraphics[width=\textwidth]{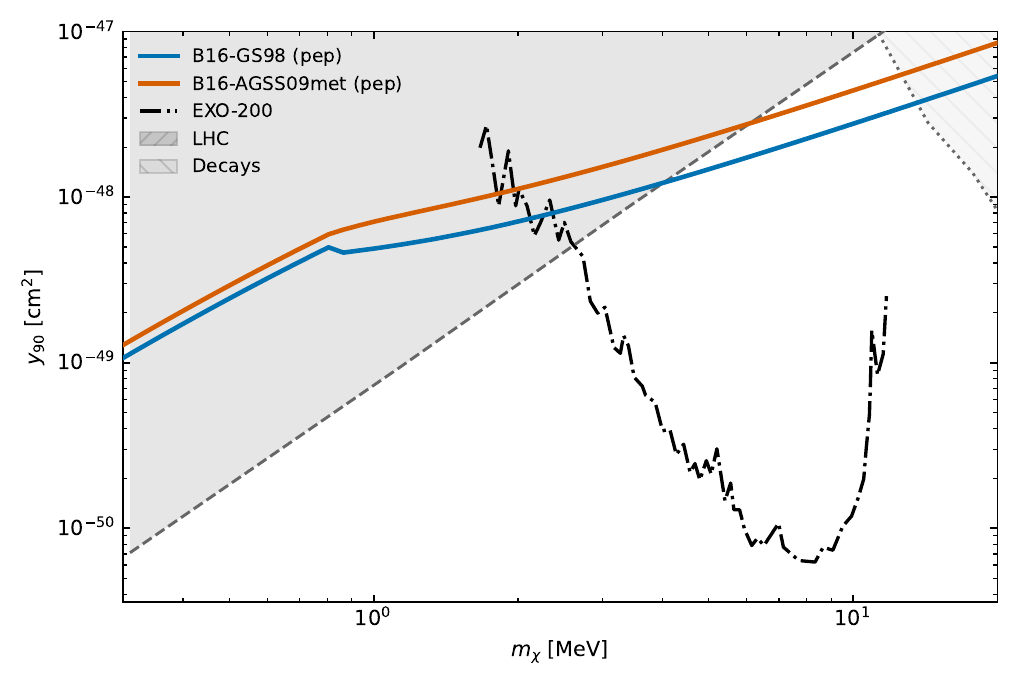}
\caption{
Main 90\% credible upper limits on charged-current fermionic DM absorption, shown as a bound on
$y = m_\chi^2/(4\pi\Lambda^4)$ as a function of the DM mass.
Curves correspond to our baseline ``pep-normalized'' operator mapping, in which the effective nuclear response is calibrated to tabulated pep $\nu_e$ capture cross sections for each target.
Results are shown for the two B16 SSM realizations (B16--GS98 and B16--AGSS09met), which bracket the dominant metallicity-driven flux systematic.
For context we overlay representative external constraints, including the EXO-200 absorption limit~\citep{EXO-200:2022adi} and the collider/decay bounds as compiled in Refs.~\citep{Dror:2020FDM_PRL,Dror:2020FDM_JHEP}.
}
\label{fig:y90}
\end{figure*}

Figure~\ref{fig:y90} presents our primary constraints on charged-current absorption, shown as a 90\% credible upper bound on $y = m_\chi^2/(4\pi\Lambda^4)$ as a function of the DM mass.
The curves start at $m_\chi \gtrsim E_{\nu,\mathrm{Ga}}^{\rm th}\simeq 0.233~\mathrm{MeV}$, reflecting the kinematic threshold for charged-current capture on ${}^{71}$Ga; below threshold the absorption-induced transition is forbidden.
Once $m_\chi \gtrsim E_{\nu,\mathrm{Cl}}^{\rm th}\simeq 0.814~\mathrm{MeV}$, the chlorine channel opens and rapidly dominates the combined likelihood. The difference between the limits obtained using B16-GS98 and
B16-AGSS09met directly reflects the metallicity dependence of
the underlying solar-neutrino flux predictions: larger predicted
capture rates leave less room for additional DM-induced
contributions and therefore yield stronger upper limits.

Figure~\ref{fig:y90} also overlays representative external constraints on the charged-current absorption, including the EXO-200 charged-current absorption search~\citep{EXO-200:2022adi} and the indirect/collider limits compiled in Refs.~\citep{Dror:2020FDM_PRL,Dror:2020FDM_JHEP}.
The collider/decay curves should be regarded as benchmark interpretations: they rely on the validity of an effective field theory (EFT) description at the relevant momentum transfers, or equivalently on assumptions about the ultraviolet (UV) completion (mediator mass/width and couplings) that generates the low-energy operator.
The radiochemical limits are complementary: they reach down to the gallium threshold ($m_\chi\simeq 0.233~\mathrm{MeV}$) and, above the chlorine threshold, become driven by the precisely measured ${}^{37}$Cl rate.

\section{Discussion}
\label{sec:discussion}
Radiochemical solar-neutrino measurements provide a simple and robust ``rate-meter'' probe of fermionic DM absorption that induces charged-current nuclear transitions on nuclei. Because the observable is a time-integrated production rate, the analysis is insensitive to detector thresholds (while being set by nuclear capture thresholds) and detailed spectral modeling, and depends primarily on the capture prediction and its uncertainties. In this sense, the resulting constraints can be viewed as fully data-driven at the rate level.

To enable a direct comparison with other absorption searches and collider interpretations, we further interpret the rate-level constraints within a specific charged-current absorption benchmark.
In this framework, the sensitivity is controlled by the ${}^{71}$Ga and ${}^{37}$Cl capture thresholds, producing a characteristic transition once the chlorine channel opens at $m_\chi\simeq 0.814~\mathrm{MeV}$.
Our operator-specific limits are therefore presented as upper bounds on $y$.

Interpreting a rate constraint in terms of $y$ requires an effective absorption kernel $A_T(m_\chi)$ (Eq.~\ref{eq:AT_def}) and introduces an operator-mapping uncertainty associated with nuclear response modeling. We minimize this additional uncertainty by adopting a pep-normalized operator mapping, in which the dominant target-dependent nuclear-structure dependence is absorbed into a single effective matrix element $M_{\rm eff,T}^2$ calibrated to tabulated pep $\nu_e$ capture inputs. Residual dependence on Coulomb corrections, recoil effects, and final-state energy dependence is expected to be subdominant at the level of current uncertainties. This operator-mapping uncertainty is conceptually distinct from the solar-metallicity systematic in the SSMs, which we represent explicitly with the B16--GS98 and B16--AGSS09met realizations~\citep{B16SSM,GS98,AGSS09}.

The resulting radiochemical limits are complementary to other probes of fermionic DM absorption.
Xenon-based searches such as EXO-200 constrain the same charged-current absorption parameterization ($y$) at higher masses~\citep{EXO-200:2022adi}, while collider and decay bounds can be interpreted in the same benchmark framework~\citep{Dror:2020FDM_PRL,Dror:2020FDM_JHEP}.
Recently, Ref.~\citep{Richardson:2025nEXO} studied the prospective sensitivity of nEXO to the charged-current absorption channel, exploiting a delayed-coincidence signature from low-lying ${}^{136}\mathrm{Cs}$ isomers.

In contrast, radiochemical data reach down to the gallium threshold, provide two independent nuclear targets, and rely on decades-long integrated rates, with minimal reliance on spectral reconstruction or event-by-event selection.

A notable advantage of the radiochemical approach is its robustness relative to collider interpretations.
The collider bounds mapped onto $(m_\chi,y)$ typically rely on the validity of an EFT description at collider momentum transfers or on a specific UV completion and depend on mediator properties and momentum transfers.
By contrast, radiochemical rates probe MeV-scale kinematics in the absorption regime and do not require extrapolating collider-scale momentum transfers to low energies. They therefore provide a comparatively assumption-light benchmark for charged-current absorption, complementary to collider constraints.

\section{Conclusion}
\label{sec:conclusion}
We derive Bayesian upper limits on fermionic DM absorption using time-integrated radiochemical solar-$\nu_e$ capture-rate measurements on ${}^{37}$Cl and ${}^{71}$Ga.
Our likelihood combines the Homestake chlorine rate~\citep{Cleveland:1998} with the weighted gallium average of SAGE, GALLEX, and GNO~\citep{Altmann:2005,Abdurashitov:2009}, and marginalizes over the dominant theory uncertainties from B16 solar fluxes~\citep{B16SSM} (bracketed by GS98 and AGSS09met~\citep{GS98,AGSS09}), capture cross sections $\sigma_T(E)$ and their uncertainties~\citep{Bahcall:1996B8Spec,Bahcall:1997GaXS,Bahcall:1998Uncertainties}, and externally constrained oscillation parameters~\citep{JUNOFirstOsc,PDG2024}.

As a rate-level (model-independent) intermediate result, we constrain non-negative additive capture-like rates $R_{\chi,\mathrm{Cl}}^{\mathrm{(free)}}$ and $R_{\chi,\mathrm{Ga}}^{\mathrm{(free)}}$ (SNU) with the combined chlorine-and-gallium likelihood; above the chlorine threshold the $R_{\chi,\mathrm{Cl}}^{\mathrm{(free)}}$ limit is driven by the small fractional uncertainty of Homestake.
We then interpret the constraints in the charged-current absorption as 90\% credible upper bounds on $y= m_\chi^2/(4\pi\Lambda^4) = \kappa m_\chi^2/(4\pi)$ from the gallium threshold at $m_\chi\simeq 0.233~\mathrm{MeV}$ up to tens of MeV.
Our limits are particularly robust in this benchmark because the operator mapping is pep-normalized, \emph{i.e.}, anchored to solar $\nu_e$ capture inputs on the same targets.
These radiochemical limits provide a simple, data-driven complement to xenon-based absorption searches such as EXO-200~\citep{EXO-200:2022adi} and to collider/decay interpretations in the same benchmark (subject to EFT/UV assumptions)~\citep{Dror:2020FDM_PRL,Dror:2020FDM_JHEP}, and show that existing solar radiochemical data probe a broad and well-motivated portion of charged-current absorption parameter space.

\section*{Acknowledgements}
This work was supported by JSPS KAKENHI Grant Numbers 24H00209, 24H02237, and 24KJ0364. 


\vspace{0.2cm}
\noindent

\let\doi\relax


\end{document}